\documentclass{article}

\usepackage{microtype}
\usepackage{graphicx}
\usepackage{subcaption}
\usepackage{booktabs}
\usepackage{multirow}
\usepackage{hyperref}

\usepackage[accepted]{icml2026} 

\usepackage{amsmath}
\usepackage{amssymb}
\usepackage{mathtools}
\usepackage{amsthm}
\usepackage[capitalize,noabbrev]{cleveref}

\theoremstyle{plain}

\icmltitlerunning{Closing the Loop: PID Feedback Control for Interpretable Activation Steering in Symbolic Music Generation}

\begin{document}

\twocolumn[
  \icmltitle{Closing the Loop: PID Feedback Control for Interpretable \\
    Activation Steering in Symbolic Music Generation}

\begin{icmlauthorlist}
\icmlauthor{Ioannis Prokopiou}{yyy,comp}
\icmlauthor{Pantelis Vikatos}{comp}
\icmlauthor{Maximos Kaliakatsos-Papakostas}{hmu}
\icmlauthor{Theodoros Giannakopoulos}{dem}
\icmlauthor{Themos Stafylakis}{yyy,ath}
\end{icmlauthorlist}

\icmlaffiliation{yyy}{Athens University of Economics and Business, Athens, Greece}
\icmlaffiliation{comp}{Orfium Research, Athens, Greece}
\icmlaffiliation{hmu}{Hellenic Mediterranean University, Chania, Greece}
\icmlaffiliation{dem}{NCSR ``Demokritos'', Athens, Greece}
\icmlaffiliation{ath}{Archimedes / Athena Research Center, Greece}

\icmlcorrespondingauthor{Ioannis Prokopiou}{gian.prokopiou@aueb.gr}

  \icmlkeywords{activation steering, PID control, symbolic music generation, sparse autoencoders, controllable generation}

  \vskip 0.3in
]

\printAffiliationsAndNotice{}

\begin{abstract}
Activation steering controls generation at inference without retraining.
In symbolic music, Sparse Activation Steering via Sparse Autoencoders enables interpretable single-layer attribute control but fails in temporal smoothing: fractional steering magnitudes fall below the Top-K sparsity threshold, zeroing the intervention.
We propose PID Steering in two forms: \emph{Spatial PID} validates control-theoretic steering in the generative music domain, while \emph{Temporal PID} dynamically adjusts $\lambda(t)$ at each autoregressive step via a closed-loop controller whose integral term accumulates error to overcome the Top-K barrier.
Experiments on the Multitrack Music Transformer show that Temporal PID overcomes the Top-K threshold failure for pitch and duration steering, enabling smooth transitions with less intervention strength and 5\% lower Fr\'{e}chet Music Distance versus static SAS.
\end{abstract}

\section{Introduction}
\label{sec:intro}

Activation steering modifies internal representations at inference to control generation without retraining~\cite{turner2024activation, li2023iti, rimsky2024steering}, based on the Linear Representation Hypothesis that concepts correspond to linear directions in activation space~\cite{park2024linear, elhage2022superposition}. \citet{nguyen2026pid} proved that static steering methods act as proportional (P) controllers unable to eliminate steady-state error caused by model priors~\cite{astrom1995pid}. We extend their spatial PID framework to the temporal axis to address a challenge absent in dense steering: the Top-K sparsity barrier.

Transformer models like the Multitrack Music Transformer (MMT)~\cite{dong2023multitrack} unfold compact token representations over long horizons where high-level concepts lack direct vocabulary mappings. Sparse Activation Steering (SAS)~\cite{bayat2025sas} provides an attractive solution~\cite{dong2023multitrack, facchiano2024musicgen, narashiman2025genre, panda2024steering}: projecting activations through SAEs enables precise, entanglement-free single-layer interventions. However, SAS imposes a strict Top-K sparsity constraint where only the largest $K$ features survive re-sparsification. This creates a \emph{binary threshold problem}---absent in dense adaptive methods~\cite{vogels2025ids, kang2026directer, li2026svf}---that defeats temporal smoothing. During a gradual cosine ramp, fractional $\lambda$ values from 0 to a target value are too small to enter the top-$K$, entirely zeroing the steering signal and forcing an abrupt binary transition instead of a smooth one (\cref{fig:topk_threshold}). We propose \textbf{Proportional-Integral-Derivative (PID) Steering} for symbolic music in two forms:
\begin{enumerate}
    \item \textbf{Spatial PID} validates the layer-wise PID formulation of \citet{nguyen2026pid} on the MMT's 12 sublayers, confirming that control-theoretic predictions hold in a shallower architecture than previously studied.
    \item \textbf{Temporal PID} transposes PID control from the spatial to the temporal domain, built around a Top-K-aware error signal: the controller measures whether steered features survive re-sparsification and adapts $\lambda(t)$ accordingly at each autoregressive step. The integral term accumulates error to overcome the Top-K threshold, enabling smooth SAS steering.
\end{enumerate}

Experiments demonstrate 62--67\% less intervention and 5\% reduced FMD degradation for PID versus static SAS, with smooth transitions that preserve generation quality. Audio examples and code can be found at \url{https://giannisprokopiouorfium.github.io/music-transformer-sae/pid}.

\section{Background}
\label{sec:background}

\paragraph{Symbolic Music Generation.}
Symbolic music generation has been reshaped by Transformer architectures \cite{vaswani2017attention}, which excel at capturing the long-term temporal dependencies inherent to musical structures. However, traditional MIDI tokenizations often suffer from sequence length explosions \cite{fradet2023miditok}. The MMT~\cite{dong2023multitrack} is a 6-block decoder-only transformer (12 sublayers, 512 dimensions, 8 attention heads) that generates polyphonic symbolic music using a compact 6-tuple event representation (type, beat, position, pitch, duration, instrument). We use a publicly available checkpoint pre-trained on the Symbolic Orchestral Database (SOD)~\cite{crestel2018sod}, which establishes a coherent generative baseline (Pitch Class Entropy $H_0{=}2.974$, Scale Consistency $S_0{=}92.26\%$, Groove Consistency $G_0{=}93.05\%$).

\paragraph{Activation Steering.}
Dense methods~\cite{turner2024activation, rimsky2024steering, arditi2024refusal, rodriguez2025meantransport} compute steering vectors as the centroid difference between contrastive datasets in the residual stream and inject $\mathbf{h}^{(\ell)} \leftarrow \mathbf{h}^{(\ell)} + \alpha \cdot \mathbf{v}^{(\ell)}$ at each layer $\ell$, but suffer from feature superposition~\cite{elhage2022superposition}: in the MMT, pitch and duration vectors exhibit cosine similarity up to 0.81 (Layer~3, \cref{app:interference}), causing interference during multi-attribute steering. SAS~\cite{bayat2025sas} resolves this by training per-layer SAEs~\cite{bricken2023monosemanticity, cunningham2024sparse} to project 512-dim activations into a 4096-dim sparse space ($8{\times}$ expansion) with strict Top-K sparsity: $f(\mathbf{a}) = \text{TopK}(\text{ReLU}(\mathbf{W}_\text{enc}\mathbf{a} + \mathbf{b}_\text{enc}), K)$. A 16-configuration layer selection grid search identified Layer~10 as optimal for SAS---it provides maximum feature capacity ($K{=}128$) with the best monotonic response across both attributes (see \cref{app:layer_selection}). At inference, the steered activation for token $t$ at layer $\ell$ is:
\begin{equation}
    \tilde{\mathbf{a}}_t^\ell = \hat{\mathbf{a}}^\ell\!\left(\sigma\!\left(f(\mathbf{a}_t^\ell) + \lambda \cdot \mathbf{v}\right)\right) + \Delta
    \label{eq:sas}
\end{equation}
where $\sigma$ is Top-K ReLU re-sparsification, $\hat{\mathbf{a}}^\ell(\cdot)$ is the SAE decoder, and $\Delta := \mathbf{a}_t^\ell - \hat{\mathbf{a}}^\ell(f(\mathbf{a}_t^\ell))$ is a reconstruction correction term that preserves information lost by the autoencoder. When $\lambda < 1$, the added features $\lambda \cdot \mathbf{v}$ have insufficient magnitude to survive $\sigma$, causing the steering signal to vanish.

\paragraph{PID Control for Steering.}
\citet{nguyen2026pid} model the layer-wise forward pass as a discrete-time dynamical system and prove that static steering is a P-controller with guaranteed steady-state error. Their PID formulation computes a dynamic steering vector $\mathbf{u}(k)$ at each layer $k$:
\begin{equation}
    \mathbf{u}(k) = K_p \mathbf{e}(k) + K_i \textstyle\sum_{j=0}^{k-1}\mathbf{e}(j) + K_d (\mathbf{e}(k) - \mathbf{e}(k{-}1))
    \label{eq:pid_spatial}
\end{equation}
where $\mathbf{e}(k) = \boldsymbol{\mu}_\text{target}(k) - \boldsymbol{\mu}_\text{source}(k)$ is the layer-wise error signal. The I~term accumulates past errors to remove residual bias; the D~term damps overshoot~\cite{astrom1995pid}. Recent adaptive methods---IDS~\cite{vogels2025ids}, DIRECTER~\cite{kang2026directer}, SVF~\cite{li2026svf}, SMITIN~\cite{luo2025smitin}---operate in dense settings where attenuated signals persist. Our temporal PID addresses the sparse threshold barrier by accumulating error until $\lambda(t)$ breaches the Top-K boundary.

\section{Method}
\label{sec:method}

We apply PID control to symbolic music steering in two forms: \emph{Spatial PID} validates the layer-wise formulation of \citet{nguyen2026pid} on the MMT's dense residual stream, confirming its predictions transfer to a shallower architecture, and \emph{Temporal PID} transposes the controller to the autoregressive time axis to solve the SAS smoothing failure.

\subsection{Spatial PID: Adapting to Symbolic Music}
\label{sec:spatial_method}

We apply \cref{eq:pid_spatial} across MMT's 12 sublayers using DiffMean vectors with an all-to-all injection strategy (\cref{app:injection}), computing the steering vector sequentially. The key challenge is the MMT's shallow depth---12 layers versus 32+ in language models~\cite{nguyen2026pid}---giving the integral term fewer steps to accumulate corrections, requiring proportionally higher $K_i$ (see~\cref{sec:spatial_results}).

\subsection{Temporal PID: Solving the Sparse Threshold Problem}
\label{sec:temporal_method}

\begin{figure}[t]
  \centering
  \includegraphics[width=\columnwidth]{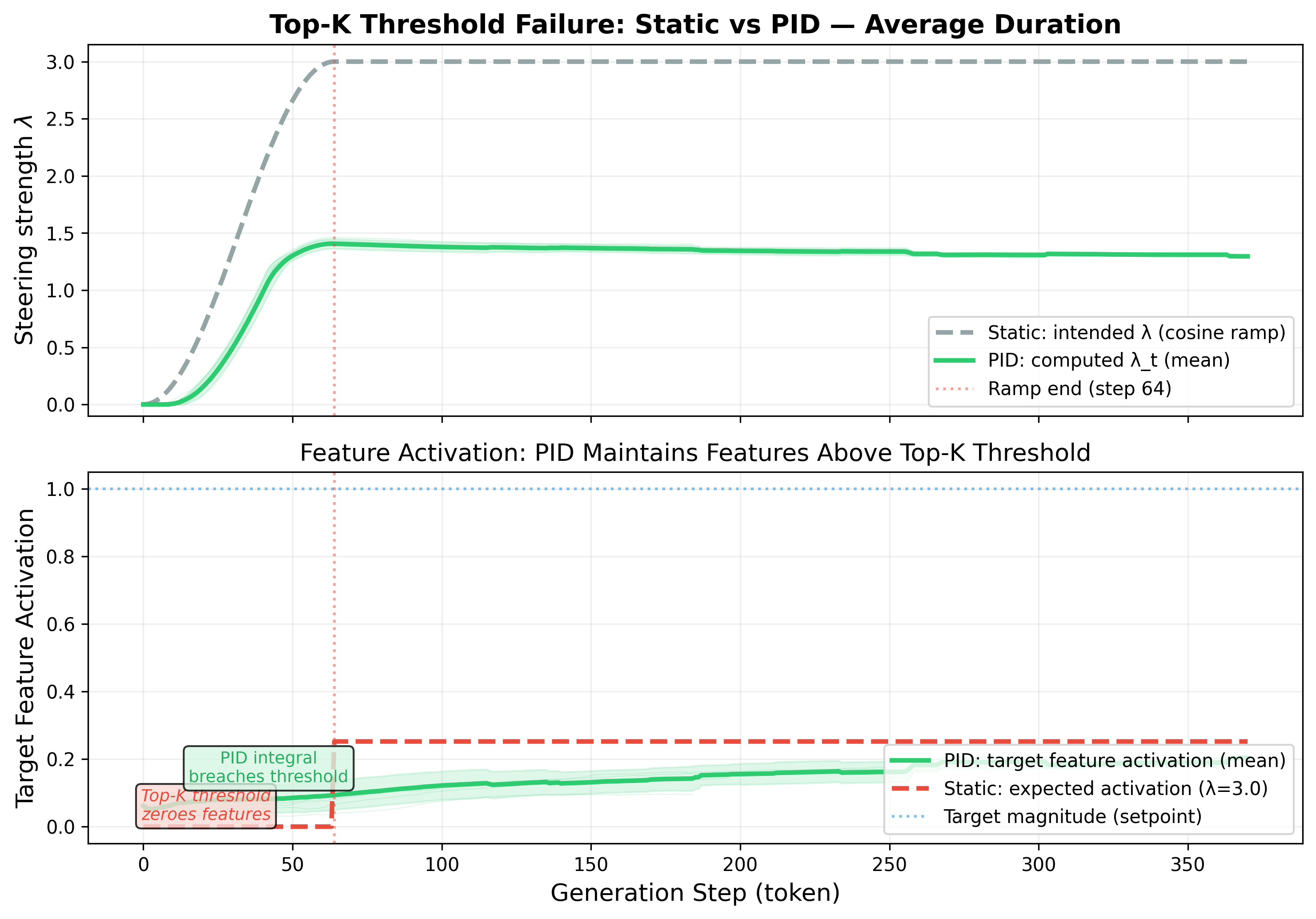}
  \caption{
    \textbf{The Top-K threshold failure.} \emph{Top:} Static cosine ramp vs.\ PID's dynamic $\lambda(t)$, which settles lower.
\emph{Bottom:} Static ramping zeros target features throughout the ramp; PID's integral accumulation maintains non-zero activations from the onset.
}
  \label{fig:topk_threshold}
  \vskip -0.2in
\end{figure}

SAS interventions in the MMT occur at a single layer (Layer~10), precluding the spatial layer-to-layer PID of \cref{sec:spatial_method}. We therefore transpose the PID control variable from the \emph{spatial} domain (layer index $k$) to the \emph{temporal} domain (generation step $t$), akin to reasoning-time temporal controllers~\cite{bharadwaj2025stupid}, creating a closed-loop feedback system that operates during autoregressive decoding.

\paragraph{Error Measurement.}
At each generation step $t$, we measure the mean magnitude of the top-$N$ target features (by absolute weight in the SAS vector $\mathbf{v}$) as $\bar{f}_a(t) = \frac{1}{|\mathcal{T}|} \sum_{j \in \mathcal{T}} f(\mathbf{a}_t^\ell)_j$, where $\mathcal{T}$ is the set of $N{=}32$ feature indices with the largest $|v_j|$. These features act as a ``concept fingerprint'' to indicate whether the steering signal survived Top-K re-sparsification. The error signal is $e(t) = m^*(t) - \bar{f}_a(t)$, where $m^*(t)$ is a cosine-ramped setpoint that smoothly scales the target activation magnitude over $T_\text{ramp}$ beats:
\begin{equation}
    m^*(t) = \begin{cases}
        \frac{m_\text{target}}{2}\left(1 - \cos\!\left(\frac{\pi \cdot t}{T_\text{ramp}}\right)\right) & t < T_\text{ramp} \\[4pt]
        m_\text{target} & t \geq T_\text{ramp}
    \end{cases}
    \label{eq:setpoint}
\end{equation}

\paragraph{PID Control Law.}
The controller computes the steering magnitude at each step:
\begin{equation}
    \lambda(t) = \text{clamp}\!\Big(K_p e(t) + K_i I(t{-}1) + K_d \big(e(t) - e(t{-}1)\big)\Big)
    \label{eq:pid_temporal}
\end{equation}
where $I(t) = \text{clamp}(I(t{-}1) + e(t),\, {-}I_\text{max},\, I_\text{max})$ is the integral accumulator with anti-windup clamping~\cite{astrom1995pid}, and $\lambda(t) \in [0, \lambda_\text{max}]$. At $t{=}0$, $I(0){=}0$, $e(-1){=}0$, and $\bar{f}_a(0)$ is the unsteered activation (typically near zero for target features). The steered sparse representation becomes $\mathbf{s}(t) = f(\mathbf{a}_t^\ell) + \lambda(t) \cdot \mathbf{v}$, followed by Top-K re-sparsification and SAE decoding as in \cref{eq:sas}.

\paragraph{Intuition.}
During the ramp, $\bar{f}_a \approx 0$ creates persistent positive error that the I~term accumulates, progressively amplifying $\lambda(t)$ until the threshold is breached. The controller then settles to a \emph{just sufficient} $\lambda$, with the D~term damping overshoot at the sub-threshold--to--active transition. We chose the mean top-$N$ feature magnitude over alternatives (e.g., the fraction of surviving features) because it provides a continuous, proportional error signal: gradual magnitude changes yield smooth $\lambda$ adjustments, whereas a binary survival count would produce bang-bang control.

\paragraph{Down-Steering and Directionality.}
Steering direction is encoded in the bidirectional SAS vector $\mathbf{v} = \mathbf{v}^{+} - \mathbf{v}^{-}$ (\cref{app:vector_construction}). For down-steering, we negate: $\mathbf{v}_\text{down} = -\mathbf{v}$; the controller output $\lambda(t) \in [0, \lambda_\text{max}]$ remains non-negative in both directions. The concept fingerprint $\mathcal{T}$ is recomputed from the active vector's largest absolute components.

\paragraph{Dual-Concept Control.}
For simultaneous steering, two independent controllers use Gram-Schmidt-orthogonalized SAS vectors with an expanded $2{\times}K$ budget at the steered layer (applied equally to PID and static baselines), since the original $K$ budget systematically displaces the second concept's orthogonalized features during Top-K selection. This scaling is an SAS framework limitation; alternative SAE architectures with relaxed sparsity may alleviate it.

\begin{figure}[t]
  \centering
  \includegraphics[width=\columnwidth]{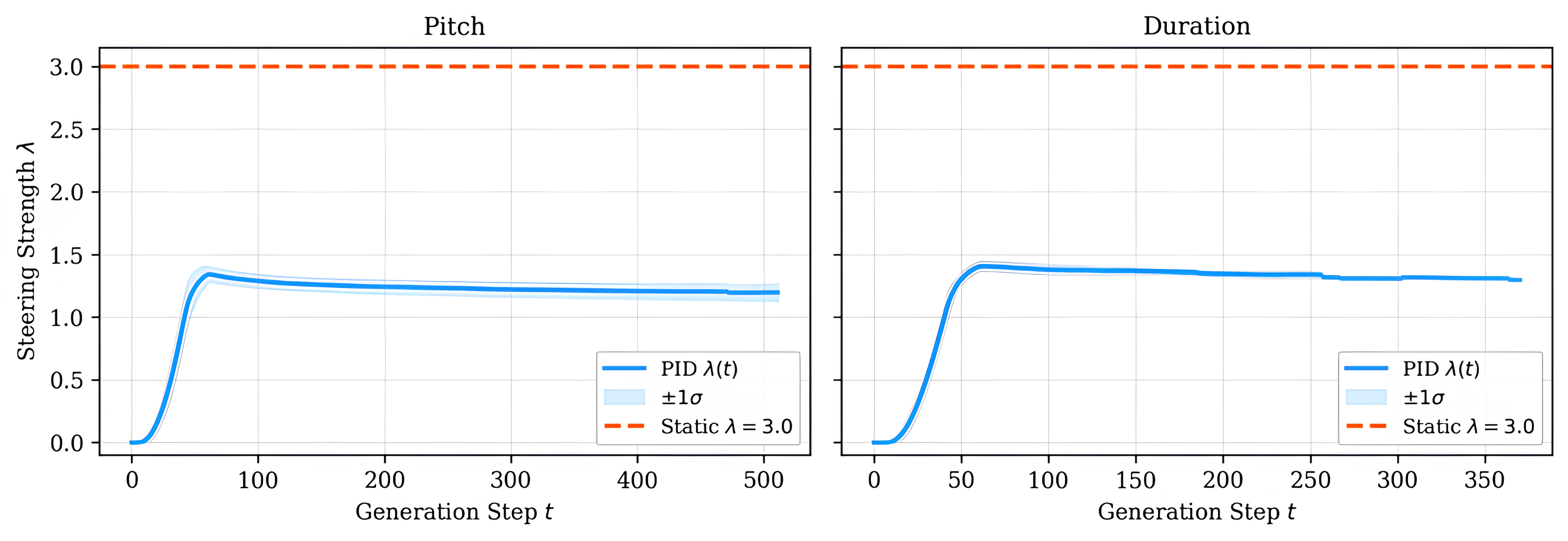}
  \caption{
    \textbf{Temporal PID $\lambda(t)$ trajectory} (steer-up, $m_\text{target}{=}2.0$). The controller overshoots briefly at threshold breach, then settles to $\lambda \approx 1.15$---62\% below static SAS's $\lambda{=}3.0$.
}
  \label{fig:hero_trajectory}
  \vskip -0.2in
\end{figure}

\section{Experiments}
\label{sec:experiments}

We evaluate on the SOD corpus using contrastive sets (1,280 samples each, \cref{app:vector_construction}) defined by the 20th/80th percentiles of pitch ($\leq$60 vs.\ $\geq$67.6 semitones) and duration ($\leq$6.5 vs.\ $\geq$14.5 ticks). SAEs are trained at each MMT layer ($512 {\to} 4096$, Top-K $K{=}128$; details in \cref{app:sae_training}). Quality degradation $\delta = |H - H_0| + \max(0, S_0 - S) + \max(0, G_0 - G)$ combines deviations from ground-truth pitch class entropy ($H_0$), scale consistency ($S_0$), and groove consistency ($G_0$); FMD uses CLaMP2 embeddings~\cite{wu2024clamp2} (details in \cref{app:metrics}). All generations use temperature 1.0, top-$k$ logit filtering (threshold 0.9), $\text{max\_seq\_len}{=}1024$, and $T_\text{ramp}{=}64$ steps.

\subsection{Spatial PID: Domain Validation}
\label{sec:spatial_results}

We validate spatial PID (\cref{eq:pid_spatial}) using DiffMean vectors across MMT's 12 sublayers ($n{=}25$, $\alpha \in \{\pm 0.5, \ldots, \pm 2.0\}$). A grid search (\cref{app:grid_search}) reveals pitch demands $8{\times}$ higher $K_i$ than duration ($0.2$ vs.\ $0.025$), reflecting stronger autoregressive priors. Error convergence replicates~\citet{nguyen2026pid}: PI/PID eliminate P-control's residual error within 12 layers. Per-$\alpha$ breakdowns appear in \cref{app:alpha_sweep}. Hook ablation across 9 configurations (\cref{app:hook_ablation}) shows all\_12 are most effective, with PID reducing $\delta$ by 17--47\% versus P-only; FMD sweeps (\cref{app:fmd_sweep}) confirm improved distributional fidelity at $\alpha{=}1.0$.

\subsection{Temporal PID: Single-Concept Steering}
\label{sec:single}

\cref{tab:single} presents results with $K_p{=}1.0$, $K_d{=}0.01$, $I_\text{max}{=}10$, $\lambda_\text{max}{=}3.0$, and concept-specific $K_i$: $0.05$ for pitch, $0.025$ for duration (a $2{\times}$ ratio, distinct from the spatial $8{\times}$ ratio from \cref{app:grid_search}). Note that the reported degradation gap between single-concept ($\delta{\approx}0.45$, $n{=}40$) and specific threshold baselines (\cref{app:threshold_baselines}, $\delta{=}0.07$, $n{=}20$) stems from sample size and subset variance. For pitch-up ($n{=}40$), PID achieves 72.65~st ($\delta{=}0.45$ [0.33, 0.62]) versus static SAS 72.30 ($\delta{=}0.64$ [0.47, 0.85]) with 62\% less intervention (avg $\lambda{=}1.15$; \cref{fig:hero_trajectory}). Negative pitch shifts $-24.0$~semitones while maintaining 98.7\% groove. Duration uses avg $\lambda \approx 1.0$---67\% less than static's $\lambda{=}3.0$. Duration-down matches static quality ($\delta{=}3.62$ vs.\ $3.37$; overlapping CIs, \cref{app:confidence_intervals}); duration-up incurs higher $\delta{=}8.45$ vs.\ $2.84$: matched-$\lambda$ analysis (\cref{app:matched_lambda}) shows static SAS at $\lambda{=}1.0$ maintains 91.3\% scale vs.\ PID's 84.7\%, indicating within-sequence $\lambda$ variation amplifies degradation for this attribute. PID's activations evolve smoothly ($\text{std}(\Delta \bar{f}_a) < 0.003$; \cref{app:smoothness}) versus static SAS's binary step. The concept fingerprint is robust to $|\mathcal{T}|$ (range 0.37~st across $N \in \{8,\!16,\!32,\!64\}$; \cref{app:fingerprint}), and PID outperforms step-function and minimal-$\lambda$ baselines (\cref{app:threshold_baselines}). Gain sweeps (\cref{app:gain_robustness}) confirm stability across $2{\times}$ perturbations (pitch 72.8--75.1~st); $T_\text{ramp}$ sensitivity (\cref{app:tramp_sensitivity}) identifies a sweet spot at $\{32, 64\}$ steps.

\begin{table}[t]
  \caption{Single-concept temporal PID steering ($\lambda_\text{max}{=}3.0$, $n{=}40$). Pitch: $K_i{=}0.05$; Duration: $K_i{=}0.025$. Static SAS uses fixed $\lambda{=}3.0$; PID uses dynamic $\lambda(t)$. Base.\ indicates the unsteered Baseline.}
  \label{tab:single}
  \begin{center}
  \begin{small}
  \begin{tabular}{llccc}
    \toprule
    Concept & Direction & PID & Static & Base.
\\
    \midrule
    \multirow{2}{*}{Pitch (semitones)} & $\uparrow$ & \textbf{72.65} & 72.30 & 68.79 \\
    & $\downarrow$ & \textbf{43.99} & 44.91 & 67.94 \\
    \multirow{2}{*}{Dur.\ (ticks)} & $\uparrow$ & 18.87 & \textbf{22.17} & 7.99 \\
    & $\downarrow$ & 4.23 & \textbf{3.35} & 7.72 \\
    \midrule
    \multicolumn{2}{l}{Pitch FMD ($\downarrow$)} & \textbf{461.9} & 487.7 & 381.5 \\
    \multicolumn{2}{l}{Duration FMD ($\downarrow$)} & \textbf{501.2} & 525.9 & 385.3 \\
    \bottomrule
  \end{tabular}
  \end{small}
  \end{center}
  \vskip -0.1in
\end{table}

\paragraph{FMD Analysis.}
PID achieves 5.3\% lower FMD than static SAS for pitch (\cref{tab:single}), as dynamic $\lambda(t)$ avoids over-steering early tokens. Note that all steered conditions increase FMD over the unsteered baseline (e.g., pitch PID 461.9 vs.\ baseline 381.5), since effective steering necessarily shifts the output distribution; PID's advantage is minimizing this drift relative to static SAS. Setpoint sweeps show $m_\text{target}{=}2.0$ is optimal for dual steering; (\cref{app:fmd_sweep}).

\paragraph{Component Ablation.}
\label{sec:ablation}
P-only control yields $\lambda_\text{avg}{=}0.664$---too conservative for the Top-K threshold. Adding I ($\lambda_\text{avg}{=}1.136$) drives $\lambda$ above threshold via error accumulation; D marginally improves settling ($\lambda_\text{avg}{=}1.158$), confirming the integral term is \emph{essential}.

\subsection{Dual-Concept Steering}
\label{sec:dual}

\cref{tab:dual} evaluates simultaneous pitch and duration steering using two independent temporal PID controllers ($m_\text{target}{=}2.0$, $n{=}20$). Per-concept adaptation manifests in magnitude (pitch settles at $\lambda{\approx}1.15$, duration at $\lambda{\approx}1.05$).

\begin{table}[t]
  \caption{Dual-concept steering ($n{=}20$). Conditioned: 16-beat prefixes, $10{\times}2$ reps. Notation: L=Low, H=High, S=Short. Full attribute values in \cref{app:conditioned_dual}.}
  \label{tab:dual}
  \begin{center}
  \begin{small}
  \begin{tabular}{lcccc}
    \toprule
    & \multicolumn{2}{c}{$\delta$ ($\downarrow$)} & \multicolumn{2}{c}{Dual\%} \\
    \cmidrule(lr){2-3} \cmidrule(lr){4-5}
    Setting & PID & Stat & PID & Stat \\
    \midrule
    Uncond.
 & \textbf{0.47} & 2.19 & 90 & \textbf{95} \\
    L/S$\to$H/L & 4.13 & \textbf{3.72} & \textbf{80} & 75 \\
    H/L$\to$L/S & 5.21 & \textbf{3.61} & \textbf{95} & 90 \\
    L/L$\to$H/S & \textbf{2.36} & 2.85 & 80 & \textbf{85} \\
    H/S$\to$L/L & \textbf{1.92} & 4.30 & 100 & 100 \\
    \bottomrule
  \end{tabular}
  \end{small}
  \end{center}
  \vskip -0.1in
\end{table}

PID achieves $4.7{\times}$ lower degradation in unconditioned steering ($\delta{=}0.47$ vs.\ $2.19$) and excels in the hardest opposing-direction conditioned case (H/S$\to$L/L: $2.2{\times}$ advantage). PID wins $\delta$ in 3 of 5 settings. In the two conditioned cases where static wins (L/S$\to$H/L, H/L$\to$L/S), PID still achieves high success rates (80--95\%), though its dynamic $\lambda$ trajectory occasionally amplifies scale degradation relative to the static baseline's uniform intervention (see \cref{app:matched_lambda}). The dual success gap (90\% vs.\ 95\% unconditioned) reflects conservative $\lambda$ occasionally under-steering duration.

\subsection{Round-Trip Steering}
\label{sec:roundtrip}

Temporal PID enables \emph{reversible steering}: steer away from a conditioned prefix, hold, then steer back---impossible with static SAS's fixed $\lambda$. Using an asymmetric three-phase schedule with 16-beat conditioned prefixes ($n{=}20$ per scenario; details in \cref{app:roundtrip}), PID outperforms a passive release baseline ($\lambda{=}0$ in Phase~3) by 8--26 percentage points (aggregate recovery: 46--74\% vs.\ 36--62\%), confirming that recovery is actively driven by closed-loop back-steering, not passive relaxation.

\section{Conclusion}
\label{sec:conclusion}

We introduced PID Steering for symbolic music: \emph{Spatial PID} validates control-theoretic steering~\cite{nguyen2026pid} in a shallow architecture, while \emph{Temporal PID} overcomes SAS's Top-K threshold failure via integral error accumulation, enabling smooth sparse steering with 62--67\% less intervention and 5\% reduced FMD degradation for pitch. Round-trip steering demonstrates reversible multi-phase trajectories that static methods cannot express, achieving 46--74\% recovery and outperforming passive release by 8--26~pp. Gain robustness sweeps confirm stability across $2{\times}$ perturbations (\cref{app:gain_robustness}), with only $+1.9\%$ marginal overhead versus static SAS (\cref{app:runtime}).

\paragraph{Limitations.}
We evaluate on a single model (MMT) and dataset (SOD); gain portability and the $2{\times}$ $K_i$ asymmetry require cross-architecture validation. Sample sizes ($n{=}40$) are modest, and perceptual validation (e.g., MUSHRA or A/B tests) is absent. Furthermore, duration-up steering degrades scale consistency to 84.7\%, which matched-$\lambda$ analysis (\cref{app:matched_lambda}) confirms is intrinsic to PID's adaptive trajectory. Future work includes adaptive gain scheduling and relaxed-sparsity SAEs like RouteSAE~\cite{shi2025routesae}.

\section*{Impact Statement}
This work advances controllable music generation. Activation steering techniques carry dual-use risks in broader settings; in symbolic music, the primary concern is unauthorized style imitation, which we mitigate by releasing only the method, not artist-specific vectors.

\section*{Acknowledgment}
This research was funded by the European Union’s Horizon Europe research and innovation programme under the AIXPERT project (Grant Agreement No. 101214389), which aims to develop an agentic, multi-layered, GenAI-powered framework for creating explainable and transparent AI systems. 

\bibliography{references}
\bibliographystyle{icml2026}

\newpage
\appendix

\section{Steering Vector Construction}
\label{app:vector_construction}

\paragraph{DiffMean Vectors.}
For each layer $\ell$, we intercept the summary activation $\mathbf{h}$ of the last fully contextualized token during a forward pass over contrastive sets. The steering vector is the centroid difference:
\begin{equation}
    \mathbf{v}^{(\ell)} = \frac{1}{N_+}\sum_{i=1}^{N_+}\mathbf{h}^{(\ell)}_{+,i} - \frac{1}{N_-}\sum_{j=1}^{N_-}\mathbf{h}^{(\ell)}_{-,j}
\end{equation}
where $N_+ = N_- = 1{,}280$ are samples from the ``High'' and ``Low'' contrastive pools extracted from the SOD corpus using the 20th/80th percentiles: average pitch ($\leq$60 vs.\ $\geq$67.6 semitones) and average duration ($\leq$6.5 vs.\ $\geq$14.5 ticks). These thresholds were validated to be stable across alternative extreme quantile choices.

\paragraph{SAS Vectors.}
Given sparse encodings of the contrastive sets ($S^+$ and $S^-$), behavior-specific features are isolated via frequency filtering with threshold $\tau{=}0.08$ (requiring a minimum 8\% activation frequency). Shared Feature Removal zeroes features active in both pools, yielding a bidirectional vector $\mathbf{v}^{(b,\ell)} = \mathbf{v}^{+} - \mathbf{v}^{-}$ that reinforces target features while suppressing opposing ones.

\section{Layer Selection for SAS}
\label{app:layer_selection}

A 16-configuration grid search evaluated SAS injection at individual layers and layer groups, balancing steering effectiveness against quality degradation~$\delta$. Key findings:

Multi-layer SAS broadcasting causes catastrophic failure: unlike DiffMean, cascading SAE reconstruction noise across layers produces total quality collapse. Layer~10 is optimal under the Adaptive~K strategy ($K$ scaling from 32 at Layer~0 to 128 at Layers~8--11), providing maximum feature capacity with the best monotonic response across both attributes, and a $48{\times}$ reduction in intervention footprint versus DiffMean (128 features vs.\ $512{\times}12$ dimensions).

\paragraph{Adaptive K Strategy.}
SAE sparsity scales linearly with depth: $K{=}32$ (Layer~0), $K{=}64$ (Layers~1--3), $K{=}96$ (Layers~4--7), $K{=}128$ (Layers~8--11). This accommodates increasing representational complexity in deeper layers.

\section{SAE Training Details}
\label{app:sae_training}

One SAE is trained per transformer layer. Architecture: $512 \to 4096$ ($8{\times}$ expansion), tied weights ($\mathbf{W}_\text{dec} = \mathbf{W}_\text{enc}^\top$), per-layer zero-mean unit-variance normalization. Training data: 640K activation vectors extracted via forward hooks from 10K tracks uniformly sampled from the SOD. Optimizer: Adam, LR $= 10^{-4}$. Loss:
\begin{equation}
    \mathcal{L} = \text{MSE}(\mathbf{a}, \hat{\mathbf{a}}) + \lambda_{L_1} \cdot \mathbb{E}[|f(\mathbf{a})|], \quad \lambda_{L_1} = 10^{-3}
\end{equation}

\section{Feature Interference Analysis}
\label{app:interference}

In the dense residual stream, pitch and duration DiffMean vectors exhibit an average absolute cosine similarity of \textbf{0.49} across all 12 layers, peaking at \textbf{0.81 in Layer~3}. This severe entanglement causes destructive interference during multi-attribute steering, requiring geometric decoupling via GSO ($\mathbf{v}^{\perp}_d = \mathbf{v}_d - \text{proj}_{\mathbf{v}_p}\mathbf{v}_d$). In the sparse SAE space, pitch and duration vectors have an average cosine similarity of $-0.27$ (anti-correlated) with 51.8\% shared feature overlap (958/4096 features), ranging from 26.5\% at Layer~6 to 75.0\% at Layer~3. For dual SAS steering, we apply Gram-Schmidt orthogonalization in sparse space and expand the Top-K budget to $2{\times}K$ to prevent feature competition.

\section{Spatial PID Gain Grid Search}
\label{app:grid_search}

We sweep $K_p \in \{0.5, 0.75, 1.0, 1.25, 1.5\}$, $K_i \in \{0.0, 0.025, 0.05, 0.10, 0.15, 0.20\}$, $K_d \in \{0.0, 0.01, 0.025, 0.05, 0.10\}$ (150 configurations per concept, $n{=}25$, $\alpha{=}1.0$). The objective function is $\text{Score} = |\Delta_\text{attr}| / (1 + \delta)$, balancing steering magnitude against quality degradation.

\begin{table}[h]
  \caption{Spatial PID optimal gains per concept. Score combines steering effectiveness and quality degradation.}
  \label{tab:grid_search}
  \begin{center}
  \begin{small}
  \begin{tabular}{lccccc}
    \toprule
    Concept & $K_p$ & $K_i$ & $K_d$ & Score & $\delta$@$\alpha{=}1$ \\
    \midrule
    Pitch & 1.5 & 0.2 & 0.01 & 32.86 & 7.91 \\
    Duration & 1.25 & 0.025 & 0.01 & 8.82 & 5.37 \\
    \bottomrule
  \end{tabular}
  \end{small}
  \end{center}
\end{table}

Pitch requires $8{\times}$ higher $K_i$ than duration. The model's autoregressive priors strongly resist pitch register deviations, requiring aggressive integral accumulation. Duration is inherently more responsive---even $K_i{=}0.025$ produces strong effects with conservative overshoot.

\section{Injection Strategy Comparison}
\label{app:injection}

For DiffMean, we compared three broadcast strategies: \textbf{All-to-All} (layer-specific vectors at all layers), \textbf{One-to-All} (single layer's vector broadcast), and \textbf{Some-to-Some} (targeting layer groups, e.g., deep layers 8--11). All-to-All proved optimal---distributed dense embeddings require multi-layer reinforcement---and is used for all DiffMean experiments. SAS uses single-layer injection at Layer~10 only.

\section{Hook Ablation Study}
\label{app:hook_ablation}

\begin{table}[h]
  \caption{Spatial PID hook ablation averaged over 2 concepts $\times$ 2 $\alpha$ values.
$|\Delta|$: steering effectiveness; Deg: quality degradation.}
  \label{tab:hook_ablation}
  \begin{center}
  \begin{small}
  \begin{tabular}{lccccc}
    \toprule
    Config & Layers & P $|\Delta|$ & PID $|\Delta|$ & P Deg & PID Deg \\
    \midrule
    all\_12 & 12 & 39.4 & 39.5 & 3.21 & \textbf{2.65} \\
    mid\_deep & 8 & 39.3 & 39.8 & 3.62 & 3.07 \\
    deep\_only & 4 & 41.3 & 41.3 & 4.46 & 3.19 \\
    attn\_only & 6 & 36.8 & 38.1 & 3.77 & 3.59 
\\
    ff\_only & 6 & 27.7 & 26.1 & 5.16 & 5.60 \\
    shallow\_only & 4 & 9.7 & 12.3 & 7.15 & 7.49 \\
    mid\_only & 4 & 27.7 & 32.4 & 9.61 & 8.07 \\
    layer\_10 & 1 & 28.8 & 33.4 & 6.23 & 3.32 \\
    layer\_11 & 1 & 36.8 & 33.2 & 6.42 & 4.45 \\
    \bottomrule
  \end{tabular}
  \end{small}
  \end{center}
\end{table}

Key observations: (1) all\_12 and mid\_deep achieve the best PID degradation (2.65 and 3.07); (2) deep\_only shows the largest PID improvement (28\% $\delta$ reduction vs.\ P-only); (3) single-layer configurations (layer\_10, layer\_11) show PID's advantage is most pronounced at concentrated injection loci (47\% $\delta$ reduction at layer\_10).

\section{FMD Sweep Across $\alpha$}
\label{app:fmd_sweep}

\begin{table}[h]
  \caption{FMD across steering magnitudes ($n{=}40$ per condition).}
  \label{tab:fmd_sweep}
  \begin{center}
  \begin{small}
  \begin{tabular}{lcccc}
    \toprule
    & \multicolumn{4}{c}{$\alpha$} \\
    \cmidrule(lr){2-5}
    Method & 0.5 & 1.0 & 1.5 & 2.0 \\
    \midrule
    \multicolumn{5}{l}{\textit{Pitch}} \\
    Baseline & 412.4 & 414.5 & 413.8 & 417.6 \\
    P-only & 464.3 & 490.0 & 494.9 & 516.1 \\
    PI & 480.7 & 508.9 & 512.1 & 518.4 \\
    PID & 
479.1 & 506.3 & 512.6 & 518.6 \\
    \midrule
    \multicolumn{5}{l}{\textit{Duration}} \\
    Baseline & 387.3 & 381.6 & 382.1 & 394.2 \\
    P-only & 409.4 & 471.9 & 483.2 & 474.6 \\
    PI & 404.9 & 462.9 & 480.4 & 475.4 \\
    PID & 412.4 & \textbf{456.0} & \textbf{475.0} & 474.8 \\
    \bottomrule
  \end{tabular}
  \end{small}
  \end{center}
\end{table}

For pitch, P-only achieves lower FMD since PI/PID's stronger steering shifts the distribution further from the reference corpus---a tradeoff between steering effectiveness and distributional fidelity. For duration ($K_i{=}0.025$, conservative), PID achieves both stronger steering \emph{and} lower FMD at $\alpha{=}1.0$--$1.5$, the best of both.

\section{Full Conditioned Dual Steering Results}
\label{app:conditioned_dual}

\begin{table}[h]
  \caption{Conditioned dual-concept steering across all 4 directional scenarios (10 songs $\times$ 2 repetitions per scenario).}
  \label{tab:conditioned_full}
  \begin{center}
  \begin{small}
  \begin{tabular}{lccccc}
    \toprule
    Scenario & Method & Pitch & Dur.
 & $\delta$ & Dual\% \\
    \midrule
    \multirow{2}{*}{L/S$\to$H/L} & PID & 66.95 & 6.89 & 4.13 & \textbf{80\%} \\
    & Static & 67.12 & 7.33 & \textbf{3.72} & 75\% \\
    \midrule
    \multirow{2}{*}{H/L$\to$L/S} & PID & 57.64 & 15.93 & 5.21 & \textbf{95\%} \\
    & Static & 56.33 & 16.70 & \textbf{3.61} & 90\% \\
    \midrule
    \multirow{2}{*}{L/L$\to$H/S} & PID & 64.09 & 8.95 & \textbf{2.36} & 80\% \\
    & Static & 64.32 & 8.84 & 
2.85 & \textbf{85\%} \\
    \midrule
    \multirow{2}{*}{H/S$\to$L/L} & PID & 53.06 & 7.33 & \textbf{1.92} & 100\% \\
    & Static & 47.91 & 5.67 & 4.30 & 100\% \\
    \bottomrule
  \end{tabular}
  \end{small}
  \end{center}
\end{table}

PID achieves $\delta$ closer to ground truth in 2 of 4 scenarios (H/S$\to$L/L, L/L$\to$H/S). In the most challenging opposing-direction case (H/S$\to$L/L), PID's $2.2{\times}$ degradation advantage is most pronounced.

\section{Spatial Single-Attribute $\alpha$ Sweep}
\label{app:alpha_sweep}

\begin{table}[h]
  \caption{Pitch steering across $\alpha$ values ($n{=}25$).}
  \label{tab:alpha_pitch}
  \begin{center}
  \begin{small}
  \begin{tabular}{lcccc}
    \toprule
    $\alpha$ & Method & Pitch (st) & $\delta$ & N \\
    \midrule
    \multirow{3}{*}{$-1.0$} & P & 39.17 & 5.2 & 346 \\
    & PI & 39.31 & 5.5 & 339 \\
    & PID & 39.55 & \textbf{3.6} & 282 \\
    \midrule
    \multirow{3}{*}{$1.0$} & P & 82.08 & 5.6 & 509 \\
    & PI 
 & 82.47 & 5.2 & 481 \\
    & PID & 82.90 & \textbf{5.0} & 484 \\
    \midrule
    \multirow{3}{*}{$2.0$} & P & 78.36 & 9.2 & 419 \\
    & PI & 79.22 & 9.1 & 478 \\
    & PID & 78.86 & \textbf{9.0} & 497 \\
    \bottomrule
  \end{tabular}
  \end{small}
  \end{center}
\end{table}

PID achieves the lowest degradation at every $\alpha$ for pitch, with the advantage most pronounced at $\alpha{=}{-}1.0$ (36\% reduction). At high $\alpha$ values, all controllers converge as the model saturates.

\section{Steering Metrics}
\label{app:metrics}

\paragraph{Quality Degradation.}
$\delta = |H - H_0|
+ \max(0, S_0 - S) + \max(0, G_0 - G)$ is the cumulative deviation from ground-truth SOD corpus statistics using MusPy metrics where $H$ is Pitch Class Entropy, $S$ is Scale Consistency, $G$ is Groove Consistency, and subscript 0 denotes ground-truth SOD values. The asymmetric penalty reflects musical priors: increases in scale/groove consistency above the corpus mean are benign, whereas decreases indicate structural degradation. Entropy uses absolute deviation since both extremes (uniform randomness and single-note collapse) are undesirable.

\paragraph{Steering Success.}
A generation is successful if its mean target attribute shifts in the specified direction relative to the unsteered baseline (unconditioned) or the conditioning prefix (conditioned).

\paragraph{Fr\'{e}chet Music Distance.}
FMD is computed using CLaMP2~\cite{wu2024clamp2} embeddings with MLE Gaussian estimation over 4,474 SOD reference MIDIs, following~\citet{wu2024fmd}.

\section{Threshold-Aware Baselines}
\label{app:threshold_baselines}

To assess whether simpler dynamic strategies can solve the Top-K threshold problem, we compare temporal PID against two alternatives ($n{=}20$, positive direction):

\begin{itemize}
    \item \textbf{Step}: $\lambda{=}0$ for $t < T_\text{ramp}$, then $\lambda{=}\lambda_\text{target}$ instantly.
\item \textbf{Minimal-$\lambda$}: Per-step binary search for the smallest $\lambda$ that activates at least one target feature above the Top-K threshold.
\end{itemize}

\begin{table}[h]
  \caption{Threshold-aware baseline comparison (positive direction).}
  \label{tab:threshold}
  \begin{center}
  \begin{small}
  \begin{tabular}{llccc}
    \toprule
    Concept & Method & Attr.\ & $\delta$ ($\downarrow$) & Avg $\lambda$ \\
    \midrule
    \multirow{4}{*}{Pitch} & PID & 73.13 & \textbf{0.07} & 1.13 \\
    & Step & 71.86 & 0.32 & 2.49 \\
    & Min-$\lambda$ & 69.44 & 0.30 & 0.01 \\
    & Baseline & 70.88 & 0.14 & --- \\
    \midrule
    \multirow{4}{*}{Dur.} & PID & 59.85 & 
7.02 & 1.11 \\
    & Step & 60.37 & 3.09 & 1.99 \\
    & Min-$\lambda$ & 70.54 & 0.09 & 0.01 \\
    & Baseline & 66.49 & 0.09 & --- \\
    \bottomrule
  \end{tabular}
  \end{small}
  \end{center}
\end{table}

For pitch, PID achieves the lowest $\delta$ (0.07) with the strongest steering (73.13~st), outperforming Step ($\delta{=}0.32$, $2.2{\times}$ more $\lambda$) and Minimal-$\lambda$ (insufficient steering). For duration, the Step function is competitive in quality but uses $1.8{\times}$ more intervention. Minimal-$\lambda$ barely steers at all ($\text{avg}\,\lambda{=}0.01$), confirming that threshold-aware does not equal effective.

\section{Concept Fingerprint Sensitivity}
\label{app:fingerprint}

\begin{table}[h]
  \caption{Sensitivity to fingerprint size $|\mathcal{T}|$ ($n{=}20$, positive direction).}
  \label{tab:fingerprint}
  \begin{center}
  \begin{small}
  \begin{tabular}{lccc}
    \toprule
    $|\mathcal{T}|$ & Pitch (st) & $\delta$ & Avg $\lambda$ \\
    \midrule
    \multicolumn{4}{l}{\textit{Pitch steering}} \\
    8 & 73.18 & 0.70 & 0.945 \\
    16 & 73.14 & \textbf{0.09} & 1.048 \\
    32 & 73.34 & 1.09 & 1.136 \\
    64 & 72.97 & 0.15 & 1.236 \\
    \bottomrule
  \end{tabular}
  \end{small}
  
\end{center}
\end{table}

Pitch steering is remarkably stable: all four $|\mathcal{T}|$ values produce pitch in [72.97, 73.34]~st (range 0.37~st). Smaller fingerprints require less $\lambda$ (0.945 at $N{=}8$ vs.\ 1.236 at $N{=}64$) since fewer features need to breach the threshold. Duration shows similar steering stability across $N$ (range [18.86, 19.48]). The default $N{=}32$ is a reasonable middle ground.

\section{Confidence Intervals}
\label{app:confidence_intervals}

\begin{table}[h]
  \caption{Bootstrap 95\% CIs ($n{=}40$, 10,000 resamples, positive steering).}
  \label{tab:ci}
  \begin{center}
  \begin{small}
  \begin{tabular}{lccc}
    \toprule
    Metric & PID & Static & Base.
\\
    \midrule
    \multicolumn{4}{l}{\textit{Pitch}} \\
    Attr.\ (st) & 72.65{\scriptsize$\pm$0.36} & 72.30{\scriptsize$\pm$0.24} & 68.79{\scriptsize$\pm$2.23} \\
    $\delta$ ($\downarrow$) & \textbf{0.45}{\scriptsize$\pm$0.15} & 0.64{\scriptsize$\pm$0.19} & 1.73{\scriptsize$\pm$0.68} \\
    Scale (\%) & 95.9{\scriptsize$\pm$0.72} & 95.9{\scriptsize$\pm$0.93} & 95.2{\scriptsize$\pm$1.09} \\
    Groove (\%) & 97.1{\scriptsize$\pm$0.54} & 97.8{\scriptsize$\pm$0.32} & 93.8{\scriptsize$\pm$0.95} \\
    \midrule
    \multicolumn{4}{l}{\textit{Duration ($K_i{=}0.025$, positive)}} \\
    Attr.\ (ticks) & 18.87{\scriptsize$\pm$0.54} & 22.17{\scriptsize$\pm$0.86} & 7.99{\scriptsize$\pm$1.00} \\
    $\delta$ ($\downarrow$) & 8.45{\scriptsize$\pm$2.48} & \textbf{2.84}{\scriptsize$\pm$1.46} & 3.05{\scriptsize$\pm$1.47} \\
    Scale (\%) & 84.7{\scriptsize$\pm$2.64} 
 & 92.4{\scriptsize$\pm$2.03} & 94.1{\scriptsize$\pm$1.62} \\
    Groove (\%) & 98.4{\scriptsize$\pm$0.15} & 99.3{\scriptsize$\pm$0.07} & 93.9{\scriptsize$\pm$1.46} \\
    \bottomrule
  \end{tabular}
  \end{small}
  \end{center}
\end{table}

For pitch, PID achieves lower degradation than static SAS: $\delta{=}0.45$ [0.33, 0.62] vs.\ $0.64$ [0.47, 0.85]. For duration, duration-down PID matches static ($\delta{=}3.62$ [2.57, 4.72] vs.\ $3.37$ [2.49, 4.41]; overlapping CIs) while using 67\% less intervention. Duration-up PID incurs higher degradation ($\delta{=}8.45$ [6.03, 10.99] vs.\ $2.84$ [1.48, 4.39]), from a scale consistency drop to 84.7\%. Matched-$\lambda$ analysis (\cref{app:matched_lambda}) indicates both the SAS vector and PID's dynamic trajectory contribute to this degradation.

\section{Intervention Smoothness}
\label{app:smoothness}

We quantify smoothness via step-to-step changes in $\lambda(t)$ and concept fingerprint activations $\bar{f}_a(t)$ ($n{=}20$ per concept). PID's $\lambda(t)$ changes by at most 0.055 per step and feature activations by less than 0.022, confirming smooth transitions. Static cosine ramping, while smooth in $\lambda$ ($\text{std}(\Delta\lambda){=}0.008$--$0.017$), produces binary feature activations---zero throughout the ramp, then a single-step jump once $\lambda$ breaches the Top-K threshold (\cref{fig:topk_threshold}).

\begin{table}[h]
  \caption{Intervention smoothness metrics (mean over samples).}
  \label{tab:smoothness}
  \begin{center}
  \begin{small}
  \begin{tabular}{lcccc}
    \toprule
    Concept & $\text{std}(\Delta\lambda)$ & $\max|\Delta\lambda|$ & $\text{std}(\Delta\bar{f}_a)$ & $\max|\Delta\bar{f}_a|$ \\
    \midrule
    Pitch & 0.010 & 0.053 & 0.002 & 0.022 \\
    Duration & 0.015 & 0.055 & 0.002 & 0.018 \\
    \bottomrule
  \end{tabular}
  \end{small}
  \end{center}
\end{table}

\section{Gain Robustness}
\label{app:gain_robustness}

We sweep $K_p$ and $K_i$ independently at multiplicative factors $\{0.5, 0.75, 1.0, 1.25, 1.5, 2.0\}$ of their nominal values ($n{=}20$, positive direction).

\begin{table}[h]
  \caption{Gain robustness for pitch ($K_p{=}1.0$, $K_i{=}0.05$ nominal).}
  \label{tab:gain_pitch}
  \begin{center}
  \resizebox{\columnwidth}{!}{
  \begin{tabular}{llcccc}
    \toprule
    Sweep & Factor & $K_p$ & $K_i$ & Pitch (st) & $\delta$ \\
    \midrule
    \multirow{6}{*}{$K_p$} & 0.50 & 0.50 & 0.050 & 75.12 & 4.24 \\
    & 0.75 & 0.75 & 0.050 & 73.22 & 3.44 \\
    & 1.00 & 1.00 & 0.050 & 73.10 & 1.24 \\
    & 1.25 & 1.25 & 0.050 & 73.60 & 1.36 \\
   
 & 1.50 & 1.50 & 0.050 & 73.13 & 2.02 \\
    & 2.00 & 2.00 & 0.050 & 72.77 & 0.70 \\
    \midrule
    \multirow{6}{*}{$K_i$} & 0.50 & 1.00 & 0.025 & 73.56 & 3.75 \\
    & 0.75 & 1.00 & 0.038 & 74.00 & 2.90 \\
    & 1.00 & 1.00 & 0.050 & 73.03 & 3.54 \\
    & 1.25 & 1.00 & 0.063 & 73.33 & 0.79 \\
    & 1.50 & 1.00 & 0.075 & 72.85 & 2.37 
\\
    & 2.00 & 1.00 & 0.100 & 73.27 & 0.61 \\
    \bottomrule
  \end{tabular}}
  \end{center}
\end{table}

Pitch steering is remarkably stable: across all 12 configurations ($2{\times}$ perturbation range), pitch remains in [72.77, 75.12]~st (range 2.35~st). $\delta$ varies more (0.61--4.24) due to the sensitivity of quality metrics to small distributional shifts, but the controller functions across all tested gains.

\begin{table}[h]
  \caption{Duration Gain robustness ($K_p{=}1.0$, $K_i{=}0.025$).}
  \label{tab:gain_dur}
  \begin{center}
  \resizebox{\columnwidth}{!}{
  \begin{tabular}{llcccc}
    \toprule
    Sweep & Factor & $K_p$ & $K_i$ & Dur.\ (ticks) & $\delta$ \\
    \midrule
    \multirow{6}{*}{$K_p$} & 0.50 & 0.50 & 0.025 & 15.13 & 2.74 \\
    & 0.75 & 0.75 & 0.025 & 18.72 & 5.72 \\
    & 1.00 & 1.00 & 0.025 & 19.15 & 9.75 \\
    & 1.25 & 1.25 & 0.025 & 19.15 & 6.21 \\
    & 
1.50 & 1.50 & 0.025 & 19.08 & 7.13 \\
    & 2.00 & 2.00 & 0.025 & 19.85 & 7.63 \\
    \midrule
    \multirow{6}{*}{$K_i$} & 0.50 & 1.00 & 0.013 & 18.70 & 6.24 \\
    & 0.75 & 1.00 & 0.019 & 18.98 & 8.47 \\
    & 1.00 & 1.00 & 0.025 & 18.93 & 8.72 \\
    & 1.25 & 1.00 & 0.031 & 19.12 & 9.18 \\
    & 1.50 & 1.00 & 0.038 & 18.72 & 8.86 \\
  
    & 2.00 & 1.00 & 0.050 & 19.52 & 7.18 \\
    \bottomrule
  \end{tabular}}
  \end{center}
\end{table}

Duration steering shows similar robustness: duration remains in [15.13, 19.85]~ticks across all perturbations. The $K_p{=}0.5$ outlier (15.13~ticks) reflects insufficient proportional response to overcome the Top-K threshold quickly, but even this extreme perturbation produces meaningful steering above the baseline (7.99~ticks).

\section{$T_\text{ramp}$ Sensitivity}
\label{app:tramp_sensitivity}

We test PID across $T_\text{ramp} \in \{16, 32, 64, 128, 256\}$ with fixed nominal gains ($n{=}20$, positive pitch steering).

\begin{table}[h]
  \caption{$T_\text{ramp}$ sensitivity for pitch steering.}
  \label{tab:tramp}
  \begin{center}
  \begin{small}
  \begin{tabular}{cccc}
    \toprule
    $T_\text{ramp}$ & Pitch (st) & $\delta$ & Avg $\lambda$ \\
    \midrule
    16 & 70.07 & 10.17 & 1.216 \\
    32 & 73.86 & \textbf{0.82} & 1.162 \\
    64 & 73.56 & 2.13 & 1.167 \\
    128 & 73.41 & 4.40 & 1.076 \\
    256 & 71.20 & 4.34 & 0.903 \\
    \bottomrule
  \end{tabular}
  \end{small}
  
\end{center}
\end{table}

$T_\text{ramp} \in \{32, 64\}$ provides the best tradeoff: both achieve pitch $>73$~st with $\delta < 2.2$. Too-short ramps ($T_\text{ramp}{=}16$) cause aggressive integral accumulation and quality degradation ($\delta{=}10.17$); too-long ramps ($T_\text{ramp}{=}256$) delay integral buildup, yielding weaker steering (71.20~st) and lower avg~$\lambda$ (0.903). The default $T_\text{ramp}{=}64$ balances smooth onset against timely convergence.

\section{Matched-$\lambda$ Analysis for Duration-Up}
\label{app:matched_lambda}

To isolate whether duration-up scale degradation stems from the SAS vector or PID's dynamic trajectory, we run static SAS at $\lambda \in \{0.5, 1.0, 1.5, 2.0, 3.0\}$ ($n{=}40$, duration-up).

\begin{table}[h]
  \caption{Static SAS at matched $\lambda$ for duration-up. Compare with PID: avg $\lambda{\approx}1.0$, scale${=}84.7\%$, $\delta{=}8.45$.}
  \label{tab:matched_lambda}
  \begin{center}
  \begin{small}
  \begin{tabular}{ccccc}
    \toprule
    $\lambda$ & Dur.\ (ticks) & Scale (\%) & Groove (\%) & $\delta$ \\
    \midrule
    0.5 & 12.33 & 95.1 & 95.4 & 1.59 \\
    1.0 & 21.19 & 91.3 & 98.6 & 2.89 \\
    1.5 & 22.44 & 93.2 & 98.8 & 1.88 \\
    2.0 & 22.80 & 91.1 & 99.1 & 4.12 \\
 
    3.0 & 24.41 & 93.7 & 97.0 & 5.80 \\
    \bottomrule
  \end{tabular}
  \end{small}
  \end{center}
\end{table}

Static SAS at $\lambda{=}1.0$ maintains 91.3\% scale consistency versus PID's 84.7\% at the same average $\lambda$; all static configurations maintain scale ${\geq}91\%$. PID's dynamic trajectory---generating tokens at varying steering magnitudes within a single piece---amplifies scale degradation beyond what fixed intervention produces. This limitation is specific to duration-up; pitch steering does not exhibit it.

\section{Round-Trip Steering Results}
\label{app:roundtrip}

\cref{tab:roundtrip_pitch,tab:roundtrip_duration} report per-window attribute values for the round-trip experiment (\cref{sec:roundtrip}). Each scenario uses 10 extreme songs $\times$ 2 repetitions ($n{=}20$), with asymmetric phases: 96~tokens (steer away, $m{=}0.75$), 64~tokens (hold with $\lambda$ decay), 192~tokens (steer back, $m{=}1.25$).

\begin{table}[h]
  \caption{Round-trip pitch steering: per-window mean pitch (st). Recovery error is $|\text{W3} - \text{baseline W3}|$.
Release stops steering in Phase~3 ($\lambda{=}0$).}
  \label{tab:roundtrip_pitch}
  \begin{center}
  \resizebox{\columnwidth}{!}{
  \begin{tabular}{llccc}
    \toprule
    Scenario & Method & W1 & W2 & W3 \\
    \midrule
    \multirow{4}{*}{Low$\to$Up$\to$Dn}
    & Baseline & 49.40 & 53.81 & 57.67 \\
    & RT PID & 55.74 & 69.61 & 50.16 \\
    & Static 1-way & 65.08 & 73.60 & 73.81 \\
    & Release & 57.20 & 72.35 & 68.87 \\
    \midrule
    & \multicolumn{2}{l}{Rec.\ err.\ (PID 
 / Static):} & \multicolumn{2}{c}{11.4 / 15.7 st} \\
    \midrule
    \multirow{4}{*}{High$\to$Dn$\to$Up}
    & Baseline & 80.93 & 81.21 & 81.16 \\
    & RT PID & 69.34 & 44.45 & 67.75 \\
    & Static 1-way & 54.75 & 39.32 & 39.62 \\
    & Release & 70.81 & 45.85 & 58.99 \\
    \midrule
    & \multicolumn{2}{l}{Rec.\ err.\ (PID / Static):} & \multicolumn{2}{c}{11.1 / 35.8 st} \\
    \bottomrule
  \end{tabular}}
  \end{center}
\end{table}

PID reduces Window~3 recovery error by 27--69\% versus static one-way steering. Per-sample aggregate recovery yields 46\% (low$\to$up$\to$down) and 66\% (high$\to$down$\to$up), versus 36\% and 40\% for release (+10 and +26~pp). The release comparison confirms active recovery: in the high scenario, PID reaches 67.8~st versus release's 59.0~st. The low scenario shows mild overshoot (W3$=$50.2 vs.\ baseline 57.7~st), reflecting asymmetric model responsiveness to pitch-down steering.

\begin{table}[h]
  \caption{Round-trip duration steering: per-window mean duration (ticks). Release stops steering in Phase~3 ($\lambda{=}0$).}
  \label{tab:roundtrip_duration}
  \begin{center}
  \resizebox{\columnwidth}{!}{
  \begin{tabular}{llccc}
    \toprule
    Scenario & Method & W1 & W2 & W3 \\
    \midrule
    \multirow{4}{*}{Low$\to$Up$\to$Dn}
    & Baseline & 2.27 & 2.17 & 2.26 \\
    & RT PID & 5.13 & 13.22 & 5.49 \\
    & Static 1-way & 15.74 & 25.98 & 26.46 \\
    & Release & 3.73 & 8.68 & 6.72 \\
   
 \midrule
    & \multicolumn{2}{l}{Rec.\ err.\ (PID / Static):} & \multicolumn{2}{c}{3.9 / 24.7 ticks} \\
    \midrule
    \multirow{4}{*}{High$\to$Dn$\to$Up}
    & Baseline & 17.06 & 16.02 & 16.08 \\
    & RT PID & 14.37 & 7.24 & 18.69 \\
    & Static 1-way & 8.30 & 3.44 & 3.27 \\
    & Release & 13.39 & 5.97 & 6.48 \\
    \midrule
    & \multicolumn{2}{l}{Rec.\ err.\ (PID / Static):} & \multicolumn{2}{c}{7.9 / 12.1 ticks} \\
    \bottomrule
  \end{tabular}}
 
  \end{center}
\end{table}

Duration low$\to$up$\to$down shows the strongest recovery: 74\% aggregate (vs.\ 62\% release, +11~pp) with 84\% lower error than static (3.9 vs.\ 24.7~ticks). High$\to$down$\to$up is weaker (15\% vs.\ 7\% release) due to low peak deviation (9.3~ticks)---the model resists downward duration changes---but PID still recovers $2{\times}$ more than release.

\section{Piecewise-Constant $\lambda$ Analysis}
\label{app:piecewise}

We test whether holding $\lambda(t)$ fixed between PID updates (update intervals of 4, 8, 16 steps) mitigates duration-up scale degradation ($n{=}40$, duration-up steering).

\begin{table}[h]
  \caption{Piecewise-constant $\lambda$ for duration-up. $\lambda$ std: within-sequence standard deviation.}
  \label{tab:piecewise}
  \begin{center}
  \begin{small}
  \begin{tabular}{ccccc}
    \toprule
    Update interval & Dur.\ (ticks) & Scale (\%) & $\delta$ & $\lambda$ std \\
    \midrule
    1 (per-token) & 20.07 & 85.8 & 7.75 & 0.725 \\
    4 & 19.55 & 85.5 & 8.13 & 0.761 \\
    8 & 19.06 & 86.0 & 7.87 & 0.781 \\
    16 & 19.15 & 84.4 & 9.09 & 0.757 \\
    
\midrule
    Static SAS & 21.17 & 91.9 & 3.17 & 0.000 \\
    \bottomrule
  \end{tabular}
  \end{small}
  \end{center}
\end{table}

Scale consistency is invariant to update interval (84--86\% across all PID variants) and $\lambda$ standard deviation remains ${\approx}0.75$ regardless of hold duration, ruling out per-token jitter as the cause. The degradation is intrinsic to PID's integral accumulation trajectory, which creates a distribution of steering magnitudes that differs qualitatively from static SAS's uniform intervention.

\section{Runtime Overhead}
\label{app:runtime}

Temporal PID's per-token overhead relative to static SAS consists of three lightweight operations: (1)~computing the error signal $e(t)$ via a mean over $N{=}32$ pre-indexed feature activations (negligible), (2)~the PID control law---three scalar multiply-adds plus two clamping operations (\cref{eq:pid_temporal}), and (3)~updating the integral accumulator $I(t)$. Both static SAS and temporal PID share the dominant cost: SAE encoding ($512 {\to} 4096$), Top-K re-sparsification, and SAE decoding ($4096 {\to} 512$) at the steered layer, plus the reconstruction correction $\Delta$.

We benchmark per-token wall-clock time over $n{=}20$ samples (512 max tokens, single A10G GPU). Unsteered baseline generation runs at $9.11 \pm 0.02$~ms/token. Static SAS adds the shared SAE encode/decode pass, increasing cost to $9.50 \pm 0.67$~ms/token ($+4.3\%$). Temporal PID reaches $9.68 \pm 0.60$~ms/token ($+6.3\%$ vs.\ baseline), with the PID controller itself contributing only $+1.9\%$ marginal overhead beyond static SAS. In practice, autoregressive sampling and attention dominate: the PID-specific computation adds $\mathcal{O}(N + 1)$ scalar operations per token atop the $\mathcal{O}(d_\text{sparse})$ SAE pass ($d_\text{sparse}{=}4096$), making temporal adaptive control essentially free relative to static steering.

\end{document}